\documentclass[conference]{IEEEtran}
\IEEEoverridecommandlockouts
\usepackage{cite}
\usepackage{amsmath,amssymb,amsfonts}
\usepackage{algorithmic}
\usepackage{graphicx}
\usepackage{textcomp}
\usepackage{xcolor}
\def\BibTeX{{\rm B\kern-.05em{\sc i\kern-.025em b}\kern-.08em
    T\kern-.1667em\lower.7ex\hbox{E}\kern-.125emX}}
\begin{document}
\bstctlcite{IEEEexample:BSTcontrol}

\title{First Full-Event Reconstruction from Imaging Atmospheric Cherenkov Telescope Real Data with Deep Learning\\
\thanks{We gratefully acknowledge financial support from the agencies and organizations listed here: www.cta-observatory.org/consortium\_acknowledgment. This project has received funding from the \textit{European Union's Horizon 2020 research and innovation programme} under grant agreements No 653477 and No 824064, and from the Fondation Universit\'e Savoie Mont Blanc. This work has been done thanks to the facilities offered by the Univ. Savoie Mont Blanc - CNRS/IN2P3 MUST computing center and HPC resources from GENCI-IDRIS (Grant 2020-AD011011577) and computing and data processing ressources from the CNRS/IN2P3 Computing Center (Lyon - France). We gratefully acknowledge the support of the NVIDIA Corporation with the donation of one NVIDIA P6000 GPU for this research.}
}


\author{\IEEEauthorblockN{Mika\"el Jacquemont\IEEEauthorrefmark{1}, Thomas Vuillaume\IEEEauthorrefmark{1}, Alexandre Benoit\IEEEauthorrefmark{2}, Gilles Maurin\IEEEauthorrefmark{1},\\ Patrick Lambert\IEEEauthorrefmark{2} and Giovanni Lamanna\IEEEauthorrefmark{1}, for the CTA LST project}\IEEEauthorblockA{\IEEEauthorrefmark{1}Univ. Savoie Mont Blanc, CNRS, LAPP - IN2P3, Annecy, France\\Email: \{firstname.lastname\}@lapp.in2p3.fr}\IEEEauthorblockA{\IEEEauthorrefmark{2}LISTIC, Univ. Savoie Mont Blanc, Annecy, France\\Email: \{firstname.lastname\}@univ-smb.fr}}


\maketitle

\begin{abstract}
The Cherenkov Telescope Array is the future of ground-based gamma-ray astronomy. Its first prototype telescope built on-site, the Large Size Telescope 1, is currently under commissioning and taking its first scientific data. In this paper, we present for the first time the development of a full-event reconstruction based on deep convolutional neural networks and its application to real data. We show that it outperforms the standard analysis, both on simulated and on real data, thus validating the \emph{deep} approach for the CTA data analysis. This work also illustrates the difficulty of moving from simulated data to actual data.
\end{abstract}

\begin{IEEEkeywords}
deep learning, gamma astronomy, multitasking, learning bias, model generalization to real data
\end{IEEEkeywords}






\section{Introduction}

\begin{figure}[bt]
\centerline{\includegraphics[width=\linewidth]{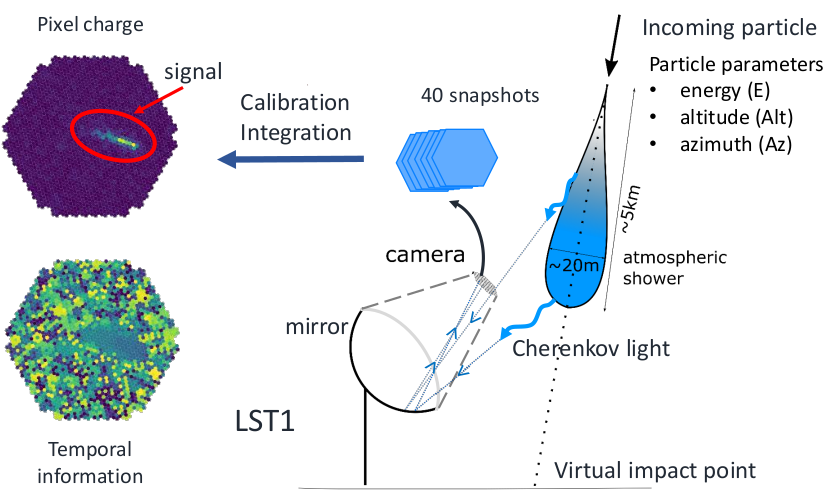}}
\caption{Large-Sized Telescope 1 (LST1) principle.}
\label{fig:iact}
\end{figure}

Astronomy, as many other domains requiring performant data analysis methods, is giving more and more attention to deep neural networks. Gamma-ray astronomy is the observation of the most energetic photons produced by violent astrophysics phenomena, such as supernova explosions, neutron star mergers or the environment of black holes.
The analysis of such phenomena will be enhanced by the Cherenkov Telescope Array (CTA), an Imaging Atmospheric Cherenkov Telescope (IACT) representing the next generation of ground-based gamma-ray telescopes, currently under construction. It will be composed of two arrays, one in each hemisphere to allow for maximal sky coverage, and will consist of tens of telescopes. The telescope prototype being built on-site on the Canary island of La Palma, the first Large Size Telescope (LST1) has started taking commissioning data recently. As shown in Fig.~\ref{fig:iact}, its method of observation, as is standard for such telescopes, relies on the observation of the Cherenkov light stimulated by the development of the atmospheric shower of particles occurring when a gamma-ray photon enters the atmosphere. This emitted light is collected by a mirror, and captured by a specific ultra-fast and very sensitive optical camera \cite{10.1117/12.2054619}. Gamma-ray-initiated showers then appear as roughly elliptical images on the camera plane.

The goal of an IACT data analysis is to reconstruct the interesting physical parameters - the particle type (gamma-ray photon vs. cosmic-ray background), its energy and incoming direction (altitude, azimuth). Several methods have been developed to achieve this event reconstruction in current generation IACTs. The most common one relies on the characterization of the moments of the image \cite{Hillas1985}. The extracted parameters are then combined with multivariate analysis methods, such as boosted decision trees or random forests \cite{Fiasson2010}. This approach, referred to as Hillas+RF in the rest of this paper, is robust, but has limited sensitivity, especially at low energies. On the other hand, state-of-the-art methods rely on template matching between the captured images and a huge dictionary, with the help of a likelihood function \cite{DeNaurois2009}. However, these methods are slow \cite{parsons2016hess}, and consequently might not be practical for CTA data analysis in their current implementation.
Indeed, to achieve its best sensitivity, CTA requires the most physically accurate analysis methods, but also methods that are computationally efficient to be able to deal with the large volume of data that will be produced (several PB per year to be analyzed again every year to benefit from calibration and analysis improvements). Deep neural networks may thus offer a solution to this challenge. 

In this paper, we present the recent developments made to address the event reconstruction problem using deep learning, and the new results obtained both on simulated and real data from LST1.
We first compare the performance of a deep multitask architecture and a widespread analysis method on the most recent production of simulated data, which takes into account the updated knowledge of the LST1. We also use an adaptation method for the simulation to better correspond to the real data. Then, we realize the first ever full-event reconstruction from IACT data with deep learning, highlighting the complexity of transferring the good performance obtained on simulations to real data.

\section{Related Work}
\label{sec:sota}

Applying deep learning techniques to the analysis of real IACT data is recent. In the High Energy Stereoscopic System (H.E.S.S.)\footnote{http://www.mpi-hd.mpg.de/hfm/HESS/pages/about/telescopes} experiment, two studies have been carried out using stereoscopic data (i.e., composed of the data from the four telescopes of H.E.S.S. I), addressing only a part of the gamma-ray event reconstruction.
Shilon \textit{et al.} \cite{shilon2019application} address the gamma/proton classification and direction regression tasks. They propose a combination of a convolutional neural network (CNN) and a Recurrent Neural Network, denoted CRNN, to handle the gamma/proton classification task. The role of the recurrent part of the model is to combine the information coming from the different telescopes. For the direction regression, they adopt a different strategy. They incorporate the data coming from the different telescopes as the different channels of a unique image, and use a shallow CNN to complete the task. Analyzing real data, they observe a decrease of the performance compared to the one on simulated data for the direction regression.
Starting from the CRNN architecture, Parsons \textit{et al.} \cite{2020EPJC...80..363P} propose to address the gamma/proton classification task by combining IACT images and standard method parameters. They observe that the obtained architecture is sensitive to the level of Night Sky Background (NSB) present in the data.

In the context of CTA, deep learning approaches have been explored, all analysis of simulated data. 
In \cite{nieto2017exploring}, Nieto \textit{et al.} probe very deep networks for gamma/proton classification from single telescope images. Reference \cite{mangano2018extracting} presents a shallower CNN to address, from stereoscopic data, gamma/proton classification, and energy and direction regression tasks. For the LST event reconstruction, the \textit{TRN-single-tel} model proposed in \cite{nieto2020adass} consists of a custom ResNet \cite{he2016deep} augmented with the Squeeze-and-Excitation attention mechanism \cite{hu2018squeeze}. Each task (i.e., gamma/proton classification, energy and direction regression) is addressed with a different model. This strategy introduces several limits, increasing the computational cost both at training and inference time while not taking into account the interdependence between the parameters to estimate.

\section{Present work}

Different from these approaches, $\gamma$-PhysNet DA presented in depth in \cite{jacquemont2021visapp} is a deep multi-task network that performs the full-event reconstruction from LST1 data with a single architecture. It is composed of a ResNet encoder and a physically inspired multi-task block. More precisely, the encoder is the convolutional part of the ResNet-56 implemented with indexed convolutions \cite{visapp19indexed}, and augmented with Dual Attention mechanism \cite{sun2020saunet}. As LST1 produces hexagonal pixel images, indexed convolutions allow applying convolution directly to the input data, avoiding additional preprocessing steps. 
In \cite{jacquemont2021visapp}, $\gamma$-PhysNet DA has proven to outperform the Hillas+RF method on all the tasks of the event reconstruction on simulated data. Also, this work highlights the appeal of the multi-task approach, thus taking into account the dependencies between the target predictions. However, these preliminary experiments were limited to simulated data. 

In this paper, we extend the work of \cite{jacquemont2021visapp} to evaluate how the performance obtained on simulations transfers to real data, highlighting the learning bias due to the difference between simulated and real data. In a first step, following the standard analysis procedure, we train $\gamma$-PhysNet DA and the Hillas+RF method on a simulated dataset (called Prod5, see Section \ref{sec:simu-data}). Compared to the data used in \cite{jacquemont2021visapp} this data set is new, and integrates the updated knowledge of the LST1 brought by the ongoing commissioning phase. Both models are then applied to the real data. However, differences remain between these updated simulations and the real data analyzed in this paper, such as the acquisition conditions or the NSB level. As emphasized in Section \ref{sec:sota}, and as we have observed in a preliminary analysis, the NSB level has a strong impact on the performance of the gamma/proton classification task with deep learning approaches. Therefore, in a second step, we measure the difference in NSB level between the simulations and the real data. We then adapt the simulations as explained in Section \ref{sec:simu_preparation} to reduce this difference in NSB level, retrain both analysis models, and reapply them to the real data.

On both the simulations and the real data the Hillas+RF method trained on standard Prod5 simulations, representing the standard approach, serves as a baseline.

\section{Data Sets}

\subsection{Simulated Data}
\label{sec:simu-data}
Ground-truth labels are impossible to obtain from IACT real data. To overcome this issue and train the reconstruction models, accurate Monte Carlo simulations have been developed to simulate the atmospheric shower development \cite{heck1998corsika} and the LST1 response \cite{Bernlohr2008}. These simulations are used to develop analysis pipelines, train machine learning algorithms, and test the reconstruction predictions.
In this work, we use the Prod5 LST1 mono trigger simulation dataset produced by the LST consortium which, for now, remains a private dataset. 
Compared to the simulations used in \cite{jacquemont2021visapp}, the Prod5 relies on the latest knowledge of the LST1 acquired during the first months of the telescope commissioning for the simulation of its response at the pointing of $20\,$\textdegree\ zenith.

\subsection{Real Data Samples of the Crab Nebula}
\label{sec:crab_data}
The real dataset is composed of two observation runs from the Crab nebula, a standard candle for gamma-ray astronomy, taken in February 2020. The observation \#2013 (
$28.9 \,$\textdegree\ zenith), denoted ON run in the rest of the paper, was taken with the telescope pointing to the gamma source direction, and the observation \#2012 (
$21.4 \,$\textdegree\ zenith), denoted OFF run, was taken with the telescope pointing to a dark patch of the sky. The runs contain respectively $10.9\,M$ and $10.4\,M$ events. As detailed later in Section \ref{sec:crab_analysis}, the OFF run allows for the estimation of the background noise that has to be subtracted from the events recorded from the source direction to obtain the gamma-ray excess.

\subsection{Data Preparation}
\label{sec:simu_preparation}
Simulated and real raw data are made of spatiotemporal data cubes of 40 samples of 1\,ns each, called waveforms. The waveforms first undergo a calibration and integration phase (the same in both cases) using lstchain v0.6.3 \cite{lstchain_adass20} in its standard configuration. This results in integrated images composed of two channels: the first, referred to as charge channel, integrating the number of photoelectrons per pixel, the second containing their mean arrival time (see Fig.~\ref{fig:iact}), thus compressing the temporal information. These images will be used as input in the later stages. 

\begin{figure}[bt]
\centerline{\includegraphics[width=0.7\linewidth]{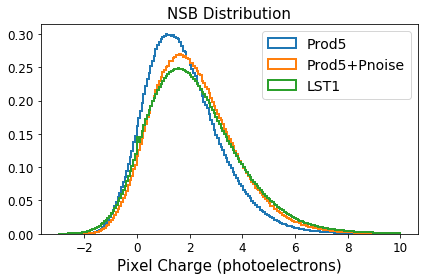}}
\caption{Distribution of NSB pixels in the simulated data (Prod5), compared to the one of the Crab data (LST1). Prod5 + Pnoise corresponds to the addition of a Poisson noise to the simulation data.}
\label{fig:noise_diff}
\end{figure}

Besides, although we benefit from high quality updated simulations to prepare the model, there still exist discrepancies between the training data and the real data produced by the LST1 that make the real data analysis challenging. The telescope pointing direction differs slightly between real data and simulations, but more importantly, the NSB level is different. Fig.~\ref{fig:noise_diff} illustrates the difference of NSB level between the simulations and the real data analyzed in this work. To overcome this issue, we add a Poisson noise to the simulated data. In the simulations, the NSB is indeed defined as a Poisson distribution. Following Raikov's theorem \cite{raikov1937decomposition}, we determine the parameter $\lambda$ of the Poisson distribution as the difference between the average noise pixel charge of the simulation and the one of the real data. The models trained on this adapted dataset are denoted \textit{+ Poisson noise} in the rest of the paper.

\subsection{Data Selection}
Finally, these images are further selected following quality cuts common in the field and required by the Hillas+RF method: the integrated signal in the charge channel must be superior to 50 photoelectrons, the ratio of the signal on the camera edges to the total signal must be lower than 20\%, and the charge channel must pass a two-level filter, called cleaning operation, requiring pixels to be above a threshold of 6 photoelectrons and to have a neighbor above 3 photoelectrons. Image samples for which none pixel survives the cleaning are discarded. 
To allow for a fair comparison, the exact same image samples are thus used with $\gamma$PhysNet DA. However, the images themselves are not cleaned from the noise in the case of $\gamma$PhysNet DA, in order to keep the information from all pixels.




After this preparation and selection phase, the test set is composed of:
\begin{itemize}
    \item for simulated data: $993\,k$ gamma-ray images from a point-like source located at the center of the field of view, $510\,k$ proton images and $773\,k$ electron images
    \item the real data as described above, for a total of $12.36\,M$ images
\end{itemize}

It is worth noticing that the models are trained on gamma rays with direction isotropically distributed within 6\,\textdegree\ of the telescope pointing, denoted diffuse gamma rays, so as to reconstruct events coming from any direction within the field of view.
In addition, no selection is applied to the training set, leaving $1.25\, M$ diffuse gamma-ray images and $836\,k$ proton images.

\section{Performance on Simulated Data\label{sec:perf-simu}}

To evaluate how both $\gamma$-PhysNet DA and the Hillas+RF method adapt to the simulated data, we compare their performance on the simulation test set, with (denoted \textit{+ Poisson noise}) and without the addition of Poisson noise for NSB level discrepancy reduction.


\subsection{Performance Metrics}
\label{sec:perf_metrics}
The \textbf{sensitivity} is a measure of the overall performance of the model. This metric is specific for gamma astronomy, and is regarded as the most relevant one because of its global aspect. It represents, per energy bin, the gamma-ray flux that an observed point-like source should emit to allow a detection with significance \cite{li1983analysis} of 5 standard deviations (denoted $\sigma$) above the background fluctuations for a 50-hour observation, with a gamma-ray excess of at least 10 events, and of at least 5\% of the residual background, per energy bin. 
In this paper, we present the sensitivity of the different methods relatively to the performance of the standard Hillas+RF method trained on standard Prod5 simulations.

The \textbf{energy resolution} represents the performance of the model for the energy reconstruction task. It is computed as the 68\% containment of the relative error of the model for the energy regression task, per energy bin.

The \textbf{angular resolution} is a measure of the performance of the model for the direction regression task. It represents the angular separation in which 68\% of the reconstructed gamma rays fall, per energy bin.

For the resolution curves, lower values indicate better performance. 
All these metrics have been computed using the \emph{pyirf} v0.4.0 package \cite{pyirf_2020_4304466} developed by the CTA community.

\subsection{Results}

\begin{figure}[bt]
\centerline{\includegraphics[width=0.8\linewidth]{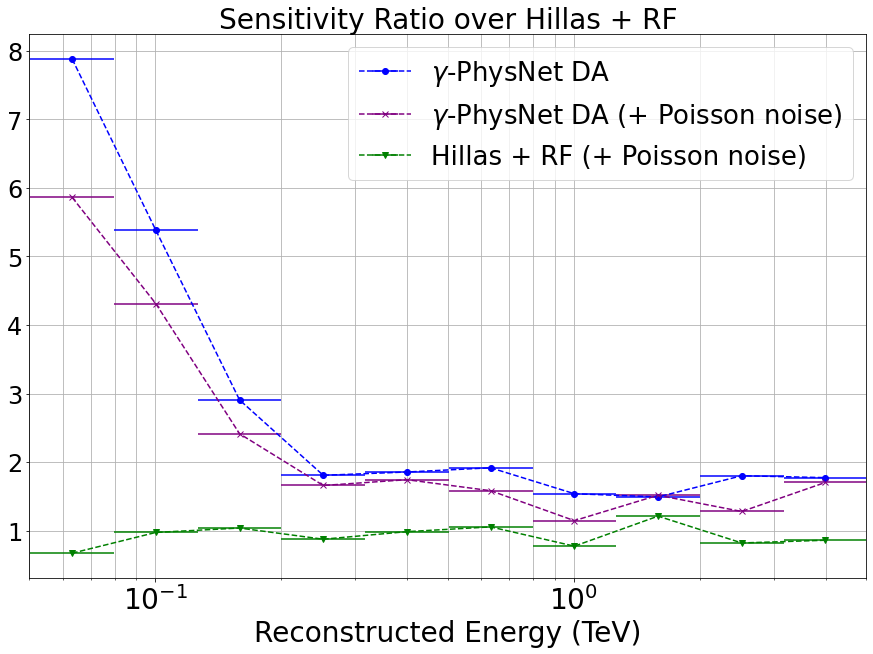}}
\caption{Ratio of sensitivity between $\gamma$-PhysNet and the Hillas+RF method.}
\label{fig:sensitivity}
\end{figure}

\begin{figure}[bt]
\centerline{\includegraphics[width=0.9\linewidth]{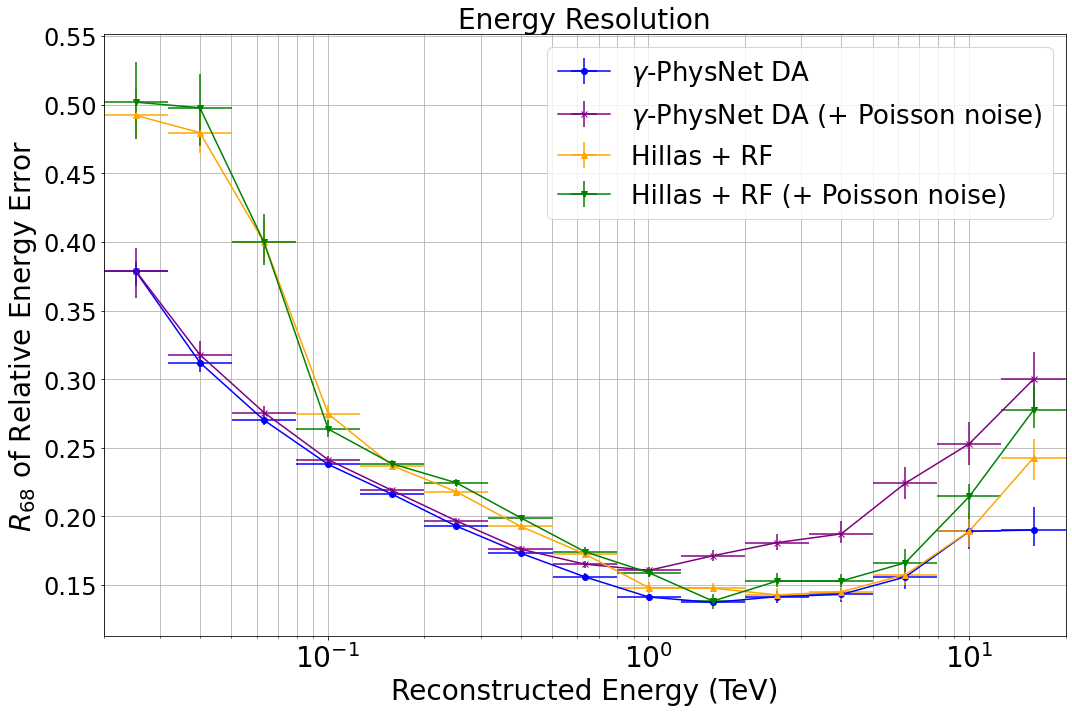}}
\caption{Energy resolution of $\gamma$-PhysNet and the Hillas+RF method (lower is better).}
\label{fig:energy_resolution}
\end{figure}

The observation of the ratio of sensitivity over the Hillas+RF method trained on standard Prod5 simulations presented in Fig.~\ref{fig:sensitivity} demonstrates the overall superiority of the deep multi-task approach on simulated data. The $\gamma$-PhysNet DA architecture improves the sensitivity compared to the Hillas+RF method by decreasing the minimal detectable flux by a factor of 1.25 to 8 below $80$\,GeV. However, we observe that the addition of Poisson noise to the data slightly degrades the performance of $\gamma$-PhysNet DA, while the Hillas+RF is less affected.
The energy resolution curves presented in Fig.~\ref{fig:energy_resolution} show that $\gamma$-PhysNet DA, the Hillas+RF and Hillas+RF \textit{+Poisson noise} obtain similar results above $100$\,GeV for the energy regression task, while $\gamma$-PhysNet DA \textit{+Poisson noise} underperforms above 2\,TeV. However, below $100$\,GeV, both $\gamma$-PhysNet DA (with and without noise) clearly outperform the standard method. These results should be considered with regards to the required energy range of the LST1 being [20\,GeV ; 3\,TeV].

\begin{figure}[bt]
\centerline{\includegraphics[width=0.9\linewidth]{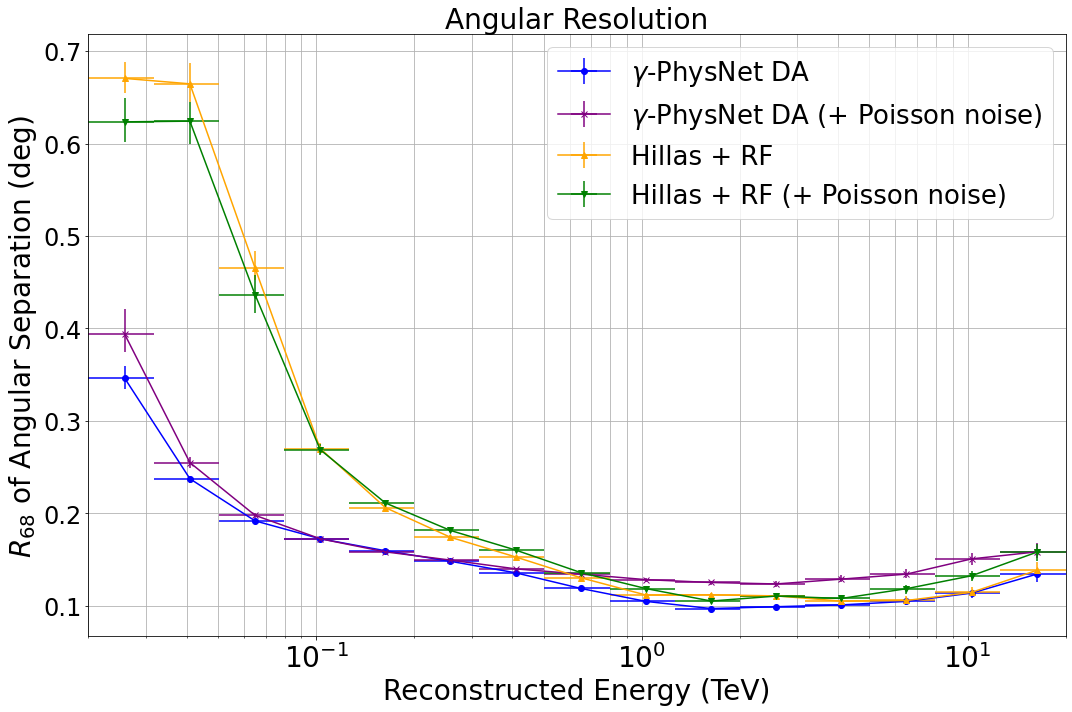}}
\caption{Angular resolution of $\gamma$-PhysNet and the Hillas+RF method (lower is better).}
\label{fig:angular_resolution}
\end{figure}

Analogously, the results for the direction regression task shown in Fig.~\ref{fig:angular_resolution} are comparable for the four models above $200$\,GeV, albeit the performance of $\gamma$-PhysNet DA slightly degrades with the addition of noise. Again, $\gamma$-PhysNet DA (with and without noise) obtains a better performance for lower energies, highlighting its greater ability to handle difficult cases, and to extend the analysis sensitivity compared to the Hillas+RF method.

From a general perspective, we observe that the performance obtained by $\gamma$-PhysNet DA slightly degrade with the addition of Poisson noise to the simulation, especially above 2\,TeV. This emphasizes the learning bias introduced into the models by the simulated data. On the contrary, the Hillas+RF method achieves similar performance with and without noise, highlighting the robustness of this method to the NSB level. 
Besides, we also observe that the deep multi-task approach constantly outperforms the Hillas+RF method for lower energies.
However, it is important to stress that transferring the good performance obtained on these simulated data to real data is challenging. 
The improvement brought by the $\gamma$-PhysNet architecture over the standard Hillas+RF method then has to be confirmed on real data.

\section{Real Data Analysis}
\label{sec:crab_analysis}

We analyze in this section the two observations of the 2020 LST1 Crab campaign described in Section \ref{sec:crab_data} with the four models: $\gamma$-PhysNet DA, $\gamma$-PhysNet DA \textit{+ Poisson noise}, the Hillas+RF method and the Hillas+RF method \textit{+ Poisson noise}. As it follows the standard approach, the analysis with the Hillas+RF method serves as a baseline for performance comparison.

\subsection{Gamma-Ray Event Detection}
\label{sec:gamma_detection}
To select in both ON and OFF runs the events detected as gamma, we define gammaness cuts, the threshold of the gamma class output of the model. 
We observe that gammaness distributions are different depending on the analysis method and on the run analyzed.
Therefore, to allow for a fair comparison of the detection performance between
the four models, we define the following procedure:
\begin{enumerate}
    \item a unique gammaness cut for all energy bins is defined for the Hillas+RF method for the ON run,
    \item for each energy bin, the gammaness cut of the three other models -- $\gamma$-PhysNet DA, $\gamma$-PhysNet DA \textit{+ Poisson noise} and Hillas+RF \textit{+ Poisson noise} -- for the ON run is tuned so that the level of background identified as signal is equivalent for all the methods,
    \item for the OFF run, the expected background (detected in the ON run by the baseline Hillas+RF) in every energy bin is first normalized to take into account the difference in acquisition duration and rate between the ON and the OFF runs. Then, the previous procedure is applied for background matching.
    \item Finally, because the acquisition conditions are different between both runs, we need to globally normalize the number of events detected in the OFF run, relying on the background level. Thanks to the previous steps, the ON-OFF normalization coefficient is similar for every analysis model.
\end{enumerate}


To ensure that the results obtained are reproducible, and not dependent on a particular gammaness cut chosen for the baseline, we repeat the analysis with 7 different cuts within the range [0.5 ; 0.8], exhibiting a common trend for the relative performance of the four models. For clarity, we only present the results obtained with a gammaness cut of 0.65.

In both runs, the background is estimated by counting the events classified as gamma rays in an area ranging from 0.32 to 0.55\,\textdegree\ from the telescope pointing direction. This area is outside the source location, even in the ON run.

\begin{figure}[bt]
\centerline{\includegraphics[width=0.95\linewidth]{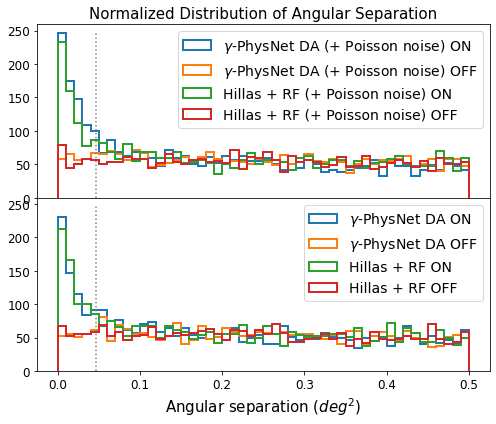}}
\caption{Gamma-ray event distribution in both observations as a function of the squared angular separation between the reconstructed and the true directions.}
\label{fig:norm_theta2}
\end{figure}

The theta squared curves shown in Fig.~\ref{fig:norm_theta2} represent the distributions of squared angular separation between the source location and the reconstructed direction. We observe that the gamma-ray event detection procedure leads to a similar level of background events (between 0.1 and 0.3 $ deg^{2}$) in the ON and the OFF runs for all the models. 

\subsection{Gamma-Ray Source Detection}

The last step of the analysis is the detection of the source itself in the ON run, based on the gamma-ray events detected in the previous step. As classically done in the field, we apply a final cut on the angular separation to the detected gamma-ray events in both runs. More precisely, events with an angular separation greater than 0.22\,\textdegree\
are discarded. Indeed, in the analysis carried out in this paper, the source location is known and events with an angular separation greater than 0.22\,\textdegree\
are more likely to be badly reconstructed.

From the surviving gamma-ray events, the following quantities are computed: 
\begin{itemize}
    \item the number of \textbf{background} events classified as gamma ray in the OFF run,
    \item the total gamma-ray \textbf{excess} in the ON run. Because analysis models are not perfect, some particles classified as gamma ray in the ON run may be background events (protons and electrons). The excess is computed by subtracting the number of background events to the number of events detected in the ON run.
    \item The statistical \textbf{significance} of the source detection.
    
\end{itemize}

As shown in Table \ref{tab:source_detection}, the Crab nebula detection by all models is clear with a significance ranging from 11.9 to 14.3\,$\sigma$ when the required value for a source detection is $5\,\sigma$. The $\gamma$-PhysNet DA models achieve similar or better performance than the Hillas+RF ones, highlighting the value of neural networks for gamma-ray astronomy.

Besides, $\gamma$-PhysNet DA \textit{+ Poisson noise} model obtains the best excess and significance, 
emphasizing the importance of simulating as accurately as possible the observation conditions. The addition of Poisson noise to the training data to reduce the NSB level difference with the real data helps lower the learning bias.

\begin{figure}[bt]
\centerline{\includegraphics[width=\linewidth]{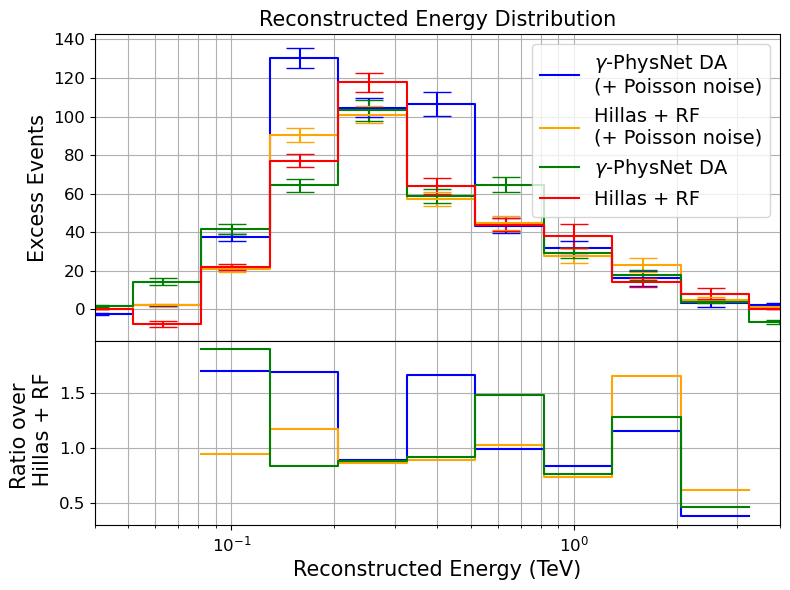}}
\caption{Excess of detected gamma rays per energy bin. The lower part of the plot represents the ratio of gamma-ray excess per energy bin detected by the models over the one detected by the baseline Hillas+RF method.}
\label{fig:gamma_excess}
\end{figure}

This is confirmed by the distribution of the excess of gamma-ray events per energy bin presented for all methods in Fig.~\ref{fig:gamma_excess}. 
As for simulated data, the gain brought by $\gamma$-PhysNet DA \textit{+ Poisson noise} architecture is more important at low energies (thanks to a greater excess for similar backgrounds).
This is especially interesting for the study of extragalactic sources and transient phenomena. Such results make it possible to exploit the full potential of the built instrument. However, these results also show a lower improvement brought by the $\gamma$-PhysNet architecture over the Hillas+RF method on real data than on simulations. This was expected as neural networks, such as $\gamma$-PhysNet, process directly the pixels of the images, while the Hillas+RF relies on extracted parameters from denoised images, thus being more robust to differences between simulations and real data. This robustness of the Hillas+RF method is underlined by the similar results obtained with or without the addition of noise to the training data.

\setlength{\tabcolsep}{10pt}
\begin{table}[bt]
\caption{Crab nebula source detection by both models}
\begin{center}
\begin{tabular}{|c|c|c|c|}
\hline
 & \textbf{Excess}& \textbf{Significance} & \textbf{Background} \\
\hline
Hillas+RF & 379 & $12.0\, \sigma$ & 308 \\
\hline
Hillas+RF & 376 & $11.9\, \sigma$ & 305 \\
\textit{+ Poisson noise} & & &  \\
\hline
$\gamma$-PhysNet DA& 395 & 12.5\, $\sigma$ & 302\\
\hline
$\gamma$-PhysNet DA& \textbf{476} & \textbf{14.3\, $\sigma$} & 317\\
\textit{+ Poisson noise} & & &  \\
\hline
\end{tabular}
\label{tab:source_detection}
\end{center}
\end{table}

\section{Conclusion}
In this paper, we propose the first ever full-event reconstruction from IACT real data with deep learning. With a deep multi-task architecture, $\gamma$-PhysNet DA, we achieve a clear detection of the Crab nebula with a statistical significance of 14.3\,$\sigma$, outperforming the standard Hillas+RF method. To obtain this result it was necessary to adapt the simulated data used to train the model by adding noise to the images, and thus taking into account the NSB level difference with the real data. 
A future work will focus on finding a more elegant solution to adapt to the NSB level that varies over the observation and over the field of view, and more generally to adapt to the data variability problem (e.g., using domain adaptation techniques).

Besides, the high-level analysis part that was beyond the focus on this study will be strengthened in a following study. In particular, an energy-dependent study of the Crab Nebula shall be made, introducing energy-dependent gammaness cuts for the baseline and deriving the spectrum of the source and comparing it to the spectrum measured by other instruments.

\bibliographystyle{IEEEtran}
\bibliography{IEEEabrv,reference}

\end{document}